# Graphether：A Two-Dimensional Oxocarbon as a Direct Wide-Gap Semiconductor with High Mechanical and Electrical Performances

Gui-Lin Zhu[1,∥], Xiao-Juan Ye[1,2,∥], and Chun-Sheng Liu[1,2,*]

[1]Key Laboratory of Radio Frequency and Micro-Nano Electronics of Jiangsu Province, College of Electronic and Optical Engineering, Nanjing University of Posts and Telecommunications, Nanjing 210023, China

[2]School of Engineering, University of British Columbia, Kelowna, BC V1V 1V7, Canada

[∥]The first two authors contributed equally to this work.

*E-mail: csliu@njupt.edu.cn

**Abstract:** Although many graphene derivatives have sizable band gaps, their electrical or mechanical properties are significantly degraded due to the low degree of $\pi$-conjugation. Besides the $\pi$-$\pi$ conjugation, there exists hyperconjugation interactions arising from the delocalization of $\sigma$ electrons. Inspired by the structural characteristics of a hyperconjugated molecule, dimethyl ether, we design a two-dimensional oxocarbon (named graphether) by the assembly of dimethyl ether molecules. Our first-principle calculations reveal the following findings: (1) Monolayer graphether possesses excellent dynamical and thermal stabilities as demonstrated by its favourable cohesive energy, absence of the soft phonon modes, and high melting point. (2) It has a direct wide-band-gap of 2.39 eV, indicating its potential applications in ultraviolet optoelectronic devices. Interestingly, the direct band gap feature is rather robust against the external strains (-10% to 10%) and stacking configurations. (3) Due to the hyperconjugative effect, graphether has the high intrinsic electron mobility. More importantly, its in-plane stiffness (459.8 N m$^{-1}$) is even larger than that of graphene. (4) The Pt(100) surface exhibits high catalytic activity for the dehydrogenation of dimethyl ether. The electrostatic repulsion serves as a driving force for the rotation and coalescence of two dehydrogenated precursors, which is favourable for the bottom-up growth of graphether. (5) Replacement of the C-C bond with an isoelectronic B-N bond can generate a stable $Pmn2_1$-BNO monolayer. Compared with monolayer hexagonal boron nitride, $Pmn2_1$-BNO has a moderate direct band gap (3.32 eV) and better mechanical property along the armchair direction.

## 1 Introduction

Graphene has emerged as a promising candidate for next-generation nanoelectronic devices because of its ultrahigh mechanical strength and excellent carrier mobility.[1] However, pristine graphene, a semimetal with a zero band gap, has an extremely low current on/off ratio, hindering its application in semiconducting devices such as field-effect transistors and logic circuits.

Several schemes have been proposed to engineer the band structures of graphene, such as spatial confinement,[2,3] sublattice-symmetry breaking,[4–6] and surface functionalization.[7–10] As shown in Table S1, covalent functionalization using hydrogen, fluorine, and oxygen provides the most effective way to obtain sizeable band gaps. Unfortunately, these surface chemical species disrupt the extended π-conjugation of graphene and thus degrade its electrical and mechanical properties.[7,11] For instance, the calculated electron mobilities of the completely hydrogenated and fluorinated graphene are only 45 and 105 cm$^2$ V$^{-1}$ s$^{-1}$, respectively.[12]

On the experimental side, the fully fluorinated or oxidized graphene films exhibit strong insulating behaviours with sheet resistance values as high as 1 TΩ sq$^{-1}$.[13–15] In addition, the Young's moduli of graphene oxide (GO) films decrease monotonically as the coverage of oxygen functional groups increases.[16,17] Therefore, it is essential to explore new two-dimensional (2D) materials with sizable band gaps while maintaining the combination of excellent electrical and mechanical properties.

Since the covalent functionalization results in the disruption of the graphene π-network, an interesting question arises whether there is a graphene derivative with σ-extended conjugation (hyperconjugation). As is well known, the hyperconjugative effects have been extensively studied in oxygen-containing molecules.[18] Furthermore, the bottom-up fabrication via on-surface assembly of molecular precursors provides a promising route to make nanomaterials. Thus it may become feasible to assemble the oxygen-containing hydrocarbons into a 2D oxocarbon with hyperconjugation. Recently, graphene sheets (Fig. 1a) have been prepared using ethylene[19] or polycyclic aromatic hydrocarbons[20,21] as precursors. On the theoretical side, some 2D materials have been designed by extending molecules into solids, such as Be$_5$C$_2$,[22] B$_2$C,[23] and Al$_2$C.[24]

To this end, two oxygen-containing compounds, i.e., epoxyethane (CH$_2$OCH$_2$) and dimethyl ether (DME, CH$_3$OCH$_3$), have been chosen. We firstly extend epoxyethane into a fully-oxidized epoxide-only phase of GO named as *Pmmn*-C$_2$O (Fig. 1b), which has also been predicted by Yan *et al*.[10] However, our results show that *Pmmn*-C$_2$O possesses low in-plane stiffness and electron mobility because of the relatively weak hyperconjugation in epoxyethane (Fig. 1b). It is possible to increase the hyperconjugative σ-electron delocalization as the number of adjacent C-H bonds increases. Clearly, the methyl group (-CH$_3$) has a greater number of C-H bonds than methylene (-CH$_2$). For instance, DME is an ether in which the oxygen atom is connected to two methyl groups. The hyperconjugation interaction is between the oxygen lone electron pairs as a donor and the σ$^*$-antibonding orbital of the C-H bond as an acceptor. This will lead to the σ (C-H) orbitals overlap between methyl groups (Fig. 1c). Therefore, the assembled 2D oxocarbon (named graphether) from DME is expected to possess the hyperconjugation effect and have considerable mechanical/electrical properties.

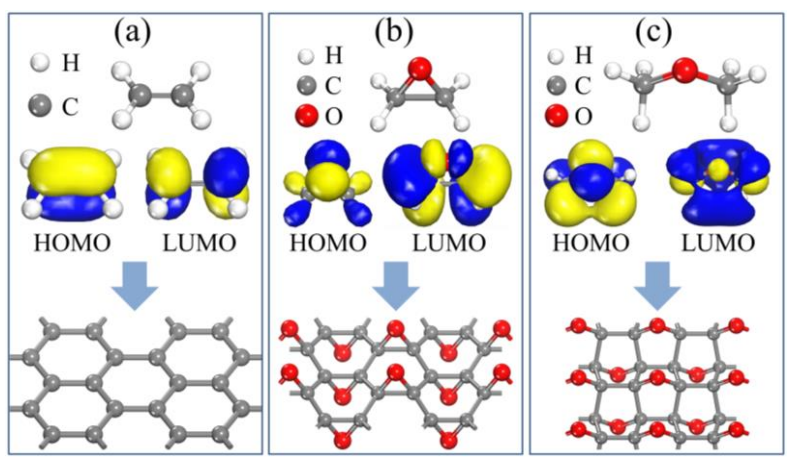

**Figure 1.** Upper panels: Geometric structures of (a) ethylene, (b) epoxyethane, and (c) DME. Middle panels: The highest occupied molecular orbitals (HOMO) and lowest unoccupied molecular orbitals (LUMO). Bottom panels: Assembly of molecules into 2D nanostructures.

In this study, we have predicted a graphether monolayer with excellent thermodynamic stabilities. Moreover, graphether has an intrinsic direct band gap of 2.39 eV which is quite robust against the strains. It exhibits anisotropic optical properties and has a strong absorption in the ultraviolet region. More interestingly, the hyperconjugation interaction leads to a high electron mobility and a large in-plane stiffness (modulus) along the armchair direction. Furthermore, we assess the feasibility for the growth of graphether on the Pt(100) surface. Finally, a stable $Pmn2_1$-BNO monolayer, an isoelectronic counterpart of graphether, has been predicted, which also shows the direct wide-band-gap semiconducting feature and attractive mechanical/electrical properties.

## 2 Computational details

All structural optimization and property calculations are performed using spin-polarized density functional theory (DFT), as implemented in the Cambridge Sequential Total Energy Package (CASTEP).[25] The generalized gradient approximation of Perdew-Burke-Ernzerhof (GGA-PBE) is used.[26] We employ the Tkatchenko-Scheffler scheme[27] to describe the van der Waals (vdW) interactions. The plane-wave kinetic energy cutoff for norm-conserving pseudopotentials[28] is set to 1230 eV, with an energy precision of $10^{-7}$ eV/atom. The $k$-points in the Brillouin zone are set to 0.02 Å$^{-1}$ and 0.01 Å$^{-1}$ spacing for the geometry optimization and electronic structure calculations, respectively. Geometry structures are fully optimized until the force is less than 0.005 eV Å$^{-1}$. A large vacuum thickness (25 Å) is used to eliminate the interaction between adjacent layers. Since the GGA usually underestimates the band gap of semiconductors, the Heyd-Scuseria-Ernzerhof (HSE06)[29] hybrid functional is adopted to correctly calculate the band structures and optical absorption coefficients. The phonon dispersion curves are computed based on the linear response method.[30] The *ab initio* molecular dynamics (AIMD) simulation in the NVT ensemble lasts for 10 ps with a time step of 1 fs. The Nosé−Hoover thermostat scheme is used to control the temperature.[31,32] The climbing-image nudged elastic band method[33,34] is used for finding minimum energy paths and energy barriers of DME dehydrogenation, diffusion, and coalescence on Pt(100).

## 3 Results and discussion

### 3.1 Structure and Stabilities

As shown in Fig. 2a, the primitive cell of graphether consists of four carbons and two oxygens with the optimized lattice parameters of $a$ = 3.614 Å and $b$ = 2.578 Å (Table 1). In comparison with $Pmmn$-$C_2O$, graphether has a comparable (smaller) value of $b$ ($a$), indicating the robust C-O interactions along the $x$ direction. The buckling of graphether, measured by the vertical distance between the bottommost O atoms and the uppermost O atoms, is slightly larger than that of $Pmmn$-$C_2O$. Moreover, the O-C-C bond angle of graphether is more close to 109.47° compared with that of $Pmmn$-$C_2O$, suggesting that the carbon site in graphether has stronger $sp^3$ hybridization.

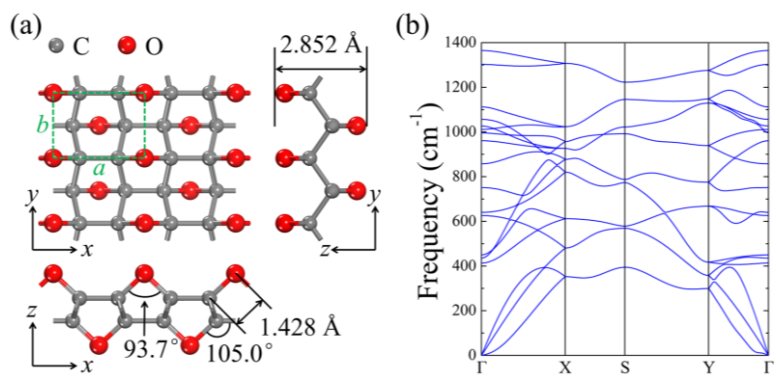

**Figure 2.** (a) Top and side views of the optimized structure of graphether. The dashed rectangle represents the unit cell. (b) Phonon dispersion of graphether.

**Table 1.** Lattice constants (*a* and *b*), buckling distance ($d_z$), O-C-C angles ($\angle$OCC), and cohesive energy ($E_{coh}$) of graphether and *Pmmn*-$C_2$O. Values in parentheses are from Ref. 35.

|  | *a* (Å) | *b* (Å) | $d_z$ (Å) | $E_{coh}$ (eV/atom) |
|---|---|---|---|---|
| graphether | 3.614 | 2.578 | 2.852 | 7.70 |
| *Pmmn*-$C_2$O | 4.443(4.44)[35] | 2.594(2.59)[35] | 2.759 | 7.72(7.90)[35] |

We first calculate the cohesive energy of graphether to evaluate its energetic stability by the expression $E_{coh} = (mE_C + nE_O - E_{total})/(m+n)$, where $E_C$ ($E_O$) and $E_{total}$ are the total energies of an isolated C (O) atom and a primitive graphether cell, respectively. *m* (*n*) is the number of C (O) atoms in the primitive cell. As shown in Table 1, the cohesive energy of graphether is very close to that of *Pmmn*-$C_2$O and comparable to that of graphene (7.91 eV/atom),[36] indicating that graphether contains robust C–O and C–C bonds.

The dynamical stability of graphether can be further assessed by calculating the phonon spectrum (Fig. 2b). The absence of imaginary phonon modes implies that graphether is dynamically stable. Particularly, the largest frequency of optical modes can reach up to 1365 cm$^{-1}$, much higher than that of black phosphorene (~450 cm$^{-1}$),[37] $MoS_2$ (~500 cm$^{-1}$),[38] and *Pmma*-CO (~1285 cm$^{-1}$),[39] indicating the robustness of covalent C-O and C-C interactions in graphether. Additionally, we perform AIMD simulations to confirm the thermal stability for graphether. As shown in Fig. 3, at temperatures below 1900 K, the structural integrity could be maintained well at the end of the simulation. However, the structure experiences serious disruption at 1900 K after only 0.14 ps, and atom positions have already been changed. Therefore, graphether may have a melting point between 1800 and 1900 K, suggesting its potential application in high temperature devices.

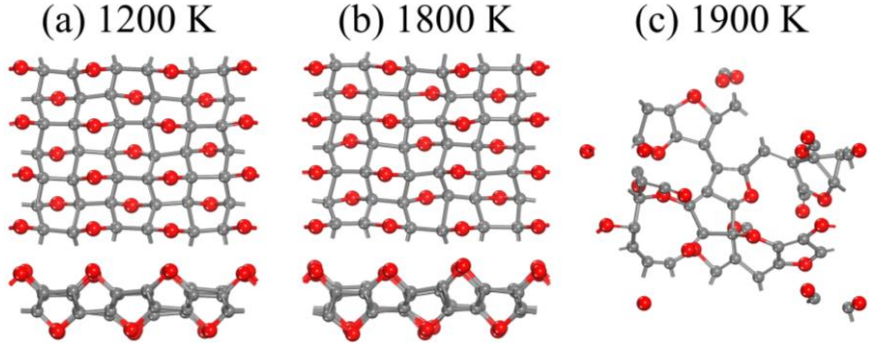

**Figure 3.** Snapshots for the final configurations of graphether at the temperatures of (a) 1200 K, (b) 1800 K, and (b) 1900 K.

**Table 2.** Calculated in-plane Young's modulus and in-plane stiffness for graphether and *Pmmn*-$C_2O$. The vdW thicknesses of graphether and *Pmmn*-$C_2O$ are 5.65 Å and 5.58 Å, respectively. Values of graphene are from refs. 40 and 41.

|  | in-plane Young's modulus (TPa) | | in-plane stiffness (N m$^{-1}$) | |
| --- | --- | --- | --- | --- |
|  | $Y_x$ | $Y_y$ | $C_x$ | $C_y$ |
| graphether | 0.81 | 0.48 | 459.8 | 273.5 |
| *Pmmn*-$C_2O$ | 0.38 | 0.48 | 212.1 | 270.1 |
| graphene | ~1.0[40] | ~1.0[40] | 342.2[41] | 342.2[41] |

### 3.2 Mechanical properties

To investigate the mechanical properties of graphether, we examine its elastic constants and in-plane Young's moduli. The computed elastic constants of graphether are $C_{11}$ = 464.22 N m$^{-1}$, $C_{22}$ = 271.53 N m$^{-1}$, $C_{12}$ = $C_{21}$ = 3.69 N m$^{-1}$, and $C_{66}$ = 116.33 N m$^{-1}$, which meet the mechanical stability criteria for the 2D orthorhombic system ($C_{11} > 0$, $C_{11}C_{22} > C_{12}^2$, $C_{66} > 0$).[42] Table 2 presents the in-plane Young's moduli and stiffnesses (see Supporting Information for details) along *x* (armchair) and *y* (zigzag) directions. Notably, graphether shows a large mechanical anisotropy with $Y_x$ ($C_x$) ~1.7 times larger than $Y_y$ ($C_y$). $Y_x$ of graphether is comparable to that of graphene, and $C_x$ is significantly larger than those of *Pmmn*-$C_2O$ and graphene, suggesting the stronger C-O or C-C bond strength along the armchair direction in graphether.

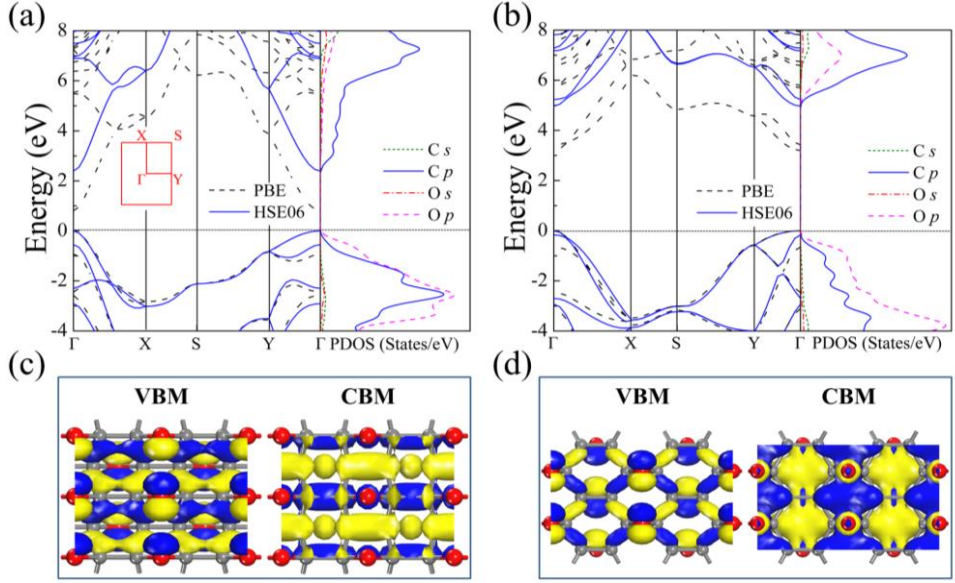

**Figure 4.** Band structure and PDOS of (a) graphether and (b) *Pmmn*-$C_2O$. Isosurfaces (0.03 e Å$^{-3}$) of electron density of the VBM and CBM at the Γ point for (c) graphether and (d) *Pmmn*-$C_2O$.

### 3.3 Electronic Structures

After confirming the excellent stability and mechanical properties of graphether, we wonder whether it has a sizeable band gap as well as high carrier mobility. As shown in Fig. 4a, graphether is a nonmagnetic semiconductor with a direct band gap of 0.81 eV (2.39 eV) calculated at the PBE (HSE06) level of theory. The band gap of graphether is evidently wider than that of monolayer black phosphorene (1.5 eV)[43] and $MoS_2$ (2.32 eV)[44] at the HSE06 level, indicating its potential applications in short-wavelength optoelectronic devices and high-power transistors. Furthermore, the partial density of states (PDOS) shows that the valence bands are predominantly composed of C (O)-2*p* orbitals. The strong overlap of C-2*p* and O-2*p* states suggests robust covalent C-O bonds. The conduction bands almost come from 2*p* orbitals of C atoms. However, the main contribution to the conduction bands of *Pmmn*-$C_2O$ is the hybridization between C-2$p$ and O-2$p$ orbitals (Fig. 4b). The participation of oxygen orbitals induces different band structures of *Pmmn*-C$_2$O.

To provide a more detailed insight into the near-band-edge states, we next examine the isosurfaces of electron density of the valence band maximum (VBM) and conduction band minimum (CBM) (Figs. 4c and 4d). Clearly, the VBM states of graphether and *Pmmn*-C$_2$O are derived from the C-C $\sigma$ states and O-2$p_y$ orbitals. For graphether, the neighboured C-C $\sigma$ states are obviously overlapped along the armchair direction, resulting from the hyperconjugation effect. Moreover, the CBM of graphether is dominated by partially delocalized $\pi$ states from the 2$p$ orbitals of carbon. Therefore, graphether may possess good intrinsic transport properties along the armchair direction. On the other hand, for *Pmmn*-C$_2$O, no overlapping $\sigma$ states are observed in its VBM state. The wave function at the CBM consists of O-2$p$ and C-2$p$ orbitals, which is localized at the hollow site of the six-membered carbon ring, restrains the formation of $\pi$ states in ethylene-like C$_2$ units. As a result, the carrier mobilities of *Pmmn*-C$_2$O would be rather small.

### 3.4 Carrier mobilities

The carrier mobility is estimated using the formula given by Lang *et al* (see Supporting Information for details).[45-47] As shown in Table 3, the electron mobilities are isotropic, but the hole mobilities show a slight direction-dependent anisotropy. In detail, the hole mobility along $y$ is 2.5 times higher than that along $x$. Additionally, in both $x$ and $y$ directions, the electron mobility is significantly larger than the hole mobility because the effective mass of electron is smaller than that of hole.

**Table 3.** Calculated DP constant ($E_{DP}$), effective mass ($m^*$), and carrier mobility ($\mu$) along $a$ and $b$ directions for graphether at 300 K.

| direction | carrier type | $E_{DP}$ (eV) | $m^*$ | $\mu$ (cm$^2$ V$^{-1}$ s$^{-1}$) |
|---|---|---|---|---|
| $x$ (armchair) | electron | -0.51 | 0.56 | 1.8×10$^3$ |
| | hole | -0.59 | 2.74 | 51 |
| $y$ (zigzag) | electron | -8.67 | 0.25 | 1.7×10$^3$ |
| | hole | -13.39 | 0.46 | 128 |

The highest electron mobility (up to 1.8×10$^3$ cm$^2$ V$^{-1}$ s$^{-1}$) is larger than those of black phosphorene (600-1580 cm$^2$ V$^{-1}$ s$^{-1}$)[48] and MoS$_2$ (200-500 cm$^2$ V$^{-1}$ s$^{-1}$).[49] In addition, graphether possesses higher electron mobility than hydrogenated and fluorinated graphene (45-105 cm$^2$ V$^{-1}$ s$^{-1}$)[12] due to the hyperconjugation effect. Moreover, the large difference between the electron and hole mobility is favourable for the electron-hole separation.

### 3.5 Optical properties

To assess the light-harvesting efficiency, we calculate the in-plane and out-of-plane absorption coefficients for incident light with the electric field (**E**) polarized along $x$ (**E**//$x$), $y$ (**E**//$y$), and $z$ (**E**//$z$) directions. As depicted in Fig. 5, the in-plane absorption coefficients exhibit large anisotropies. Monolayer graphether shows good absorption performance for **E**//$y$ in the region from 1 to 4 eV, which corresponds to the near-infrared, visible, and near-UV spectral ranges. However, the absorption edge for **E**//$x$ has a blue shift of about 3 eV compared with that of **E**//$y$. This anisotropic optical performance is beneficial to developing polarized optical sensors.

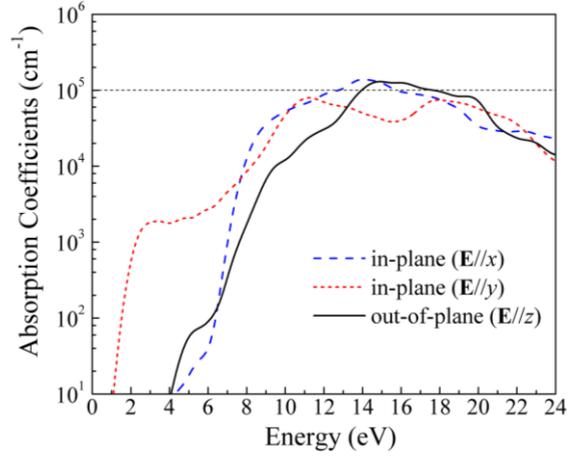

**Figure 5.** Calculated in-plane and out-of-plane absorption coefficients of graphether at the HSE06 level.

The maximum absorption coefficients for **E**//$x$ and **E**//$z$ are comparable with those of the organic perovskite solar cells.[50] Furthermore, the in-plane absorption coefficients of graphether larger than $10^5$ cm$^{-1}$ only occur around 14 eV, in contrast to those of arsenene and antimonene occurring in a wider range of 3 to 10 eV.[51,52] Thus, the high selective absorption characteristic could be used in ultraviolet light sensors and laser devices.

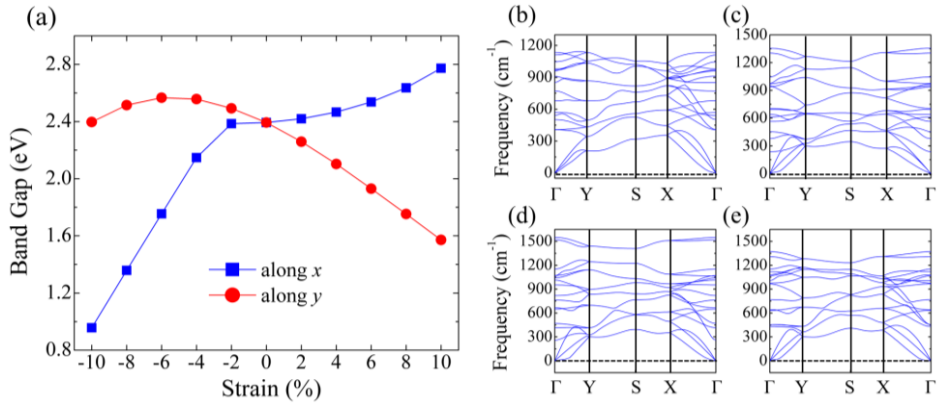

**Figure 6.** (a) Band gap of graphether under uniaxial strain calculated by the HSE06 functional. Positive and negative strains indicate the expansion and compression, respectively. The phonon spectra when the strain (b) $\varepsilon_x$ (applied in the armchair direction) = 10%, (c) $\varepsilon_y$ (applied in the zigzag direction) = 10%, (d) $\varepsilon_x$ = -4%, and (e) $\varepsilon_y$ = -4%.

## 3.6 Strain engineering

Strain engineering can serve as an effective way to control electronic properties of semiconductors.[53–55] As shown in Fig. 6a, the band gap increases monotonously with the applied strain $\varepsilon_x$ in the range from -10% to 10%, but the energy gap variation with the $y$-direction strain exhibits a parabolic trend. Especially, the band gap of graphether can be tuned over a much wider range by applying the strain along the $x$ direction. To check the stability of uniaxially strained graphether, we have computed the phonon spectra as a function of strain. When the tensile strain in the $x$ ($y$)-direction reaches 10%, no imaginary phonon frequencies are observed (Figs. 6b and 6c). However, graphether can only keep stable with the compressive strain up to 4% (Figs. 6d and 6e). Within the compressive strain range of 6-10%, graphether has apparent imaginary modes in the acoustic phonon branches (phonon spectra not shown). Therefore, graphether could withstand uniaxial strains ranging from -4% to 10%.

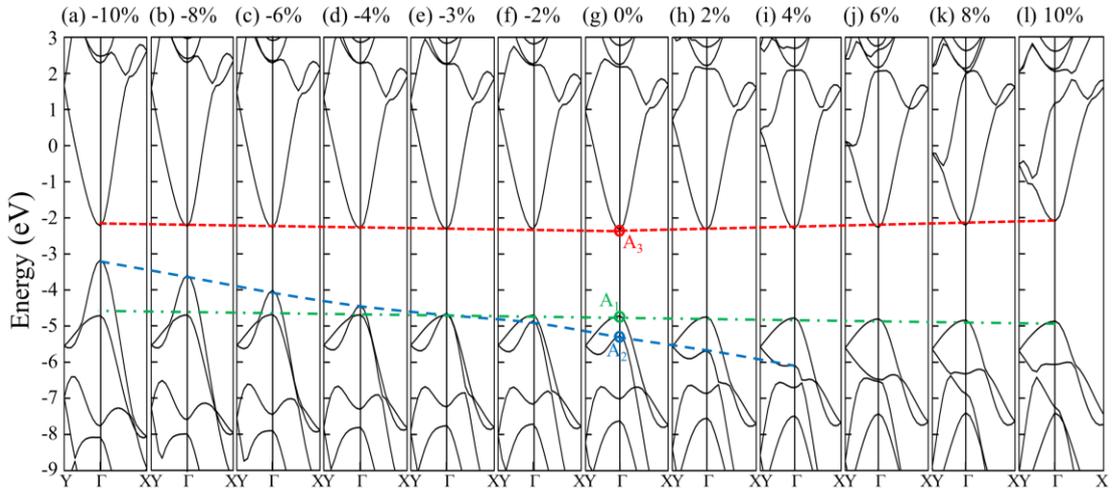

**Figure 7.** Band structure of graphether with respect to $\varepsilon_x$. The dashed lines show the energy shifts of states $A_1$, $A_2$, and $A_3$.

The energy gap evolution is dominated by the states near the band edges. Figures 7

and 8 present the strain effects on the band structure. Clearly, both the VBM and CBM are located at the Γ point under various strains. In Fig. 7, the nonlinear trend of the VBM with respect to $\varepsilon_x$ is attributed to the competition of the energies of two near-band-edge states (labelled as $A_1$ and $A_2$). With increasing compressive strain, the energy of state $A_2$ increases rapidly and is equal to that of state $A_1$ at $\varepsilon_x = -3\%$. Therefore, $A_2$ represents the VBM with the compressive strain larger than -3%. In contrast, the CBM is always represented by state $A_3$. $A_3$ is nearly independent of the uniaxial strain, and thus its energy remains flat. On the other hand, with $\varepsilon_y$ ranging from -10% to 10%, states $A_1$ and $A_3$ at Γ always represent the VBM and CBM, respectively, both of which exhibit approximate linear decreasing trends (Fig. 8).

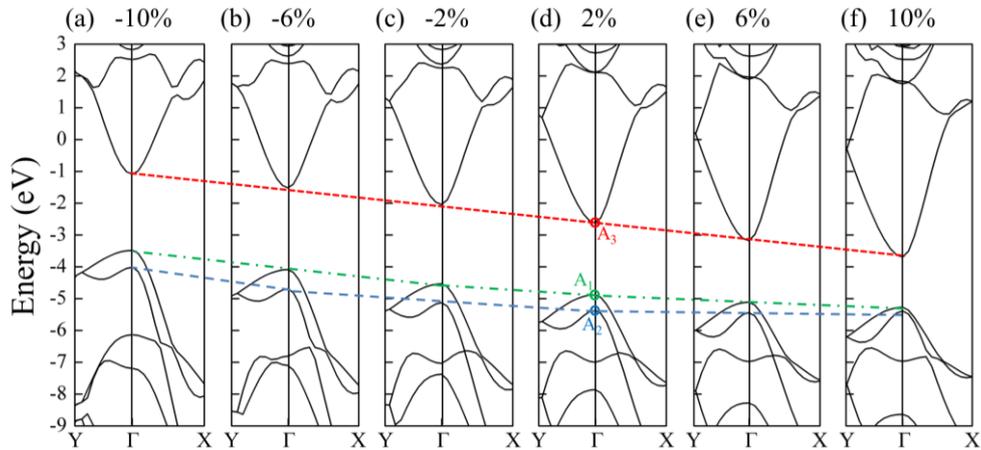

**Figure 8.** Band structure of graphether with respect to $\varepsilon_y$. The dashed lines show the energy shifts of states $A_1$, $A_2$, and $A_3$.

Based on the Heitler-London exchange energy model,[56] the energy trends of the near-band-edge states with strain have been demonstrated to be related to the characteristics of the bonding and antibonding states.[57,58] The energies of the bonding ($E_+$) and antibonding states ($E_-$) are expressed by the following formula,

$$E_{+} = 2E_{H} + \frac{Q+T}{1+S^2} \quad (1)$$

$$E_{-} = 2E_{H} + \frac{Q-T}{1-S^2} \quad (2)$$

where $E_H$ represents the energy of an unperturbed atom, $S$ is the overlap integral of the orbitals between near-neighbour atoms, $Q$ is the classical coulomb energy, and $T$ is an exchange integral term. Because $S$ is generally much less than 1, $T$ plays a decisive role in determining the bonding and antibonding energies. The $T$ is expressed by

$$T = \iint \varphi_\alpha^*(1)\varphi_\beta^*(2)[-\frac{1}{r_{\alpha 2}} - \frac{1}{r_{\beta 1}} + \frac{1}{r_{12}} + \frac{1}{R}]\varphi_\alpha(2)\varphi_\beta(1)d\tau_1\tau_2, \quad (3)$$

where the negative $-\frac{1}{r_{\alpha 2}} - \frac{1}{r_{\beta 1}}$ and positive $\frac{1}{r_{12}} + \frac{1}{R}$ terms correspond to the attractive electron-ion interaction and repulsive electron (ion)-electron (ion) interaction, respectively. For $s$ orbitals, in comparison with the electron-electron interaction, the electron-ion interaction makes a larger contribution to the exchange $T$. When the interatomic distance increases, the value of $T$ increases because the energy of the electron-ion contribution increases more quickly compared with the energy reduction of the electron-electron interaction. However, the situation is on the contrary for the bonding $p$ orbitals.

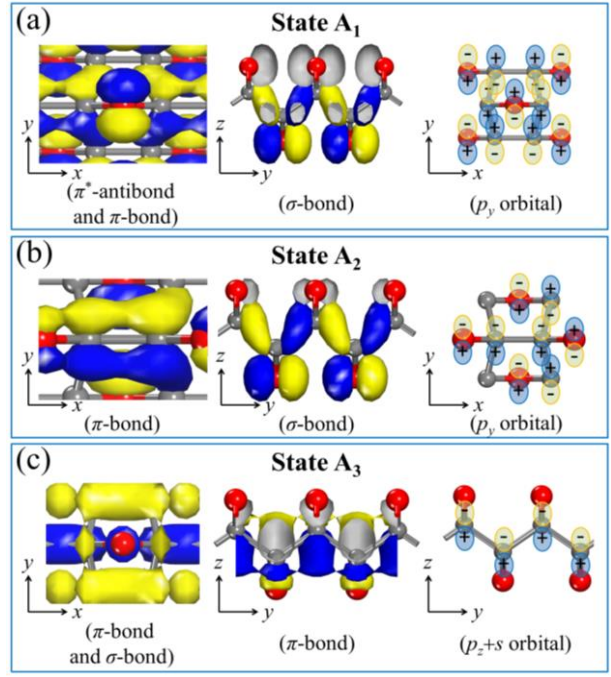

**Figure 9.** Isosurfaces (0.03 e Å$^{-3}$) of electron wave functions of (a) state $A_1$, (b) state $A_2$, and (c) state $A_3$. The rightmost panels represent the projected major orbital and sign of phase factor. The bond status and main orbitals in the horizontal axis are shown at the bottom of each subgraph.

To examine the bonding features, we analyse the electronic wave functions for states $A_1$, $A_2$, and $A_3$. States $A_1$ and $A_2$ are dominated by $2p_y$ orbitals at the sites of carbon and oxygen. State $A_3$ appears as a mixture including 55% $2p_z$ orbitals of carbon atoms and the rest from $2s$ orbitals of carbon and oxygen. In the $x$ direction, state $A_1$ (Fig. 9a) consists of C-C $\pi$-bonding and C-O $\pi^*$-antibonding, suggesting an insensitive response of its energy to the uniaxial strain. This leads to the flat curve of state $A_1$ (Fig. 7). However, $A_1$ is bonding in the $y$ direction and its energy is expected to decrease with $\varepsilon_y$ (Fig. 8). $A_2$ shows bonding in both the $x$ and $y$ directions (Fig. 9b), and thus its energy shows a decreasing trend as the tensile strain increases (Figs. 7 and 8). State $A_3$ in Fig. 9c displays C-C $\pi$-bonding and $\sigma$-bonding ($2s$ orbitals of C and O) in the $x$ direction. Its energy exhibits little change with $\varepsilon_x$ (Fig. 7) due to the opposite energy trends of bonding *s* and bonding *p* orbitals. On the other hand, state $A_3$ is π-bonding in the *y* direction and thus its energy decreases with $\varepsilon_y$. This is consistent with the curve of $A_3$ in Fig. 8.

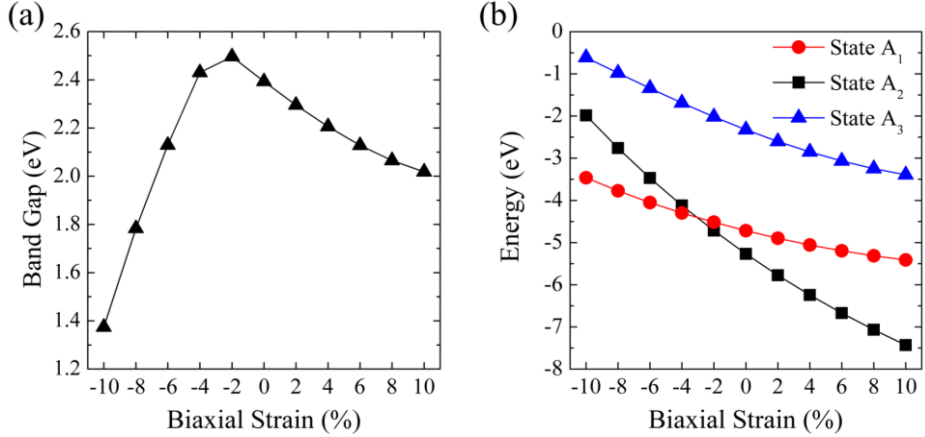

**Figure 10.** (a) Band gap of graphether and (b) energy of states $A_1$, $A_2$, and $A_3$ as a function of biaxial strain at the HSE06 level of theory.

Moreover, we calculate the band structures of graphether under the biaxial strain from -10% to 10%. In Fig. 10a, as the strain increases, the band gap initially increases from 1.37 eV to 2.50 eV and then drops to 2.08 eV. As seen from Fig. 10b, state $A_1$ represents the VBM in the strain range from -2% to 10%. When the compressive strain ranges from -2% to -10%, $A_2$ becomes the VBM because $A_2$ has a higher energy than $A_1$. Therefore, the energy crossover for states $A_1$ and $A_2$ at compression indicates that the band gap first increases and then decreases. Compared with the uniaxial strain, the biaxial strain has a lower efficiency of tunable band gap. Interestingly, both VBM and CBM are located at the Γ point, suggesting the direct band gap nature in the biaxially strained graphether.

In fact, the supporting substrate must inevitably be employed for the fabrication of practical devices using 2D materials. The lattice mismatch between the 2D layer and the substrate induces the structural deformation, which usually results in the absence of some intrinsic properties such as the direct band gap feature. For example, small strains (<2%) can trigger a direct-to-indirect gap transition in phosphorene and monolayer MoS$_2$.[57,59] In contrast, graphether can maintain the direct band gap feature within the strain range from -10% to 10%, suggesting its potential applications in flexible optoelectronic and electronic devices.

### 3.7 Bilayer structures

Layer stacking is another effective approach to tune the electronic properties of 2D materials.[60–62] We consider four possible stacking configurations of bilayer graphether, namely, AA-, AB-, AC-, AD-stacking (Fig. 11). As listed in Table S2, the *a* and *b* lattice parameters differ slightly for different stacking types. The most significant difference among the four stacking types is the interlayer distance which ranges from 2.003 Å in the AA-stacking to 3.516 Å in the AD-stacking.

The interlayer binding energy is expressed as $E_b = (E_{bi} - 2E_{mo})/S$, where $E_{bi}$ ($E_{mo}$) and $S$ are the total energy of the bilayer (monolayer) and the coupling area, respectively. The negative values of $E_b$ (Table S2) indicate that the stacking process is exothermic. Particularly, the AA-stacking is energetically most stable. For comparison, the calculated binding energy of bilayer graphene is -26.63 meV Å$^{-2}$, in agreement with those reported in the literature (-22.3~-38.43 meV Å$^{-2}$).[62] The interlayer bonding strength of AA-stacking bilayer graphether is of the same order of magnitude as that of bilayer graphene systems, indicating the vdW interaction between two graphether layers. (Fig. 11). Compared with monolayer graphether, the graphether bilayers present a small decrease (0.07-0.12 eV) in the band gaps. However, regardless of the stacking patterns, the direct nature of the band gap (at the Γ point) is still maintained.

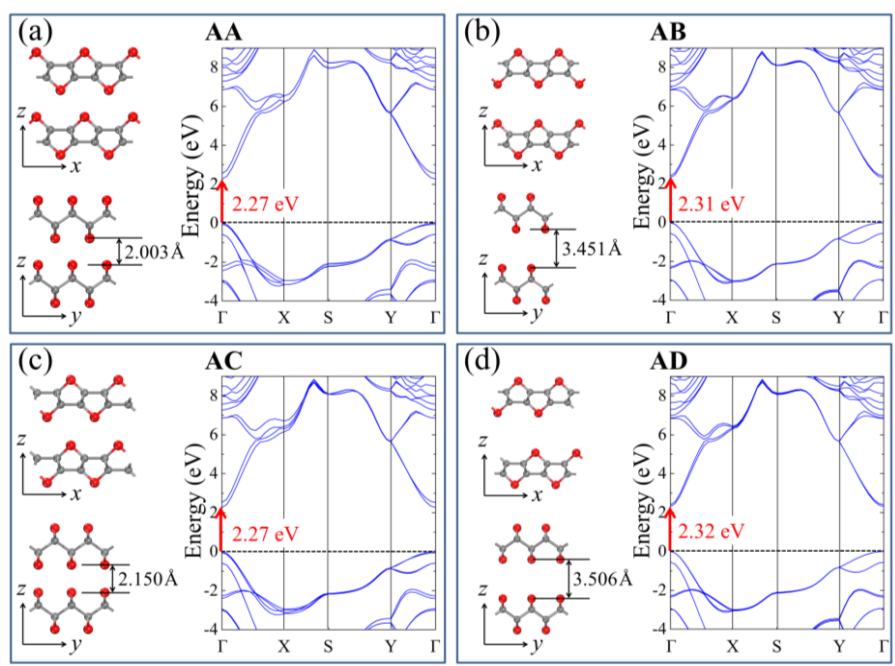

**Figure 11.** Side views of configurations and band structures (HSE06) of bilayer graphether for (a) AA-, (b) AB-, (c) AC-, and (d) AD-stacking patterns. AA: the top layer is directly stacked on the bottom layer; AB (AC): the upper layer of AA-stacking is shifted by half of a primitive cell along the $x$ ($y$) direction; AD: the top and bottom layers are mirror images of each other.

### 3.8 Potential substrate

One may wonder whether the predicted graphether can be bottom-up synthesized using DME as precursors on a suitable substrate. Herein, we consider DME deposited on Pt(100) which exhibits higher catalytic activity than other facets of Pt.[63] Previous studies have demonstrated the dehydrogenation of DME on Pt(100) prior to the C-O bond breaking.[64] Therefore, the DME assembly process is mainly composed of three stages: (1) DME adsorbed on the metal surface; (2) DME partially dehydrogenated to form the methoxymethyl intermediate ($CH_3OCH_2^*$); and (3) dehydrogenated DME rotation, diffusion and coalescence on the metal surface.

At stage (1), we investigate various adsorption sites and molecule orientations to determine the most stable configuration of DME on Pt(100). The Pt(100) surface is modelled by a four-layered 4 × 4 supercell, with the bottom two Pt layers fixed in their bulk positions during the geometry relaxation. The adsorption energy ($E_{ads}$) of DME is defined as $E_{ads} = E_{DME} + E_{Substrate} - E_{DME+Substrate}$, where $E_{DME}$, $E_{Substrate}$, and $E_{DME+Substrate}$ are the total energies of the DME molecule in the gas phase, the pristine substrate, and the DME-adsorbed substrate, respectively. As shown in Fig. 12a, the top site has the largest adsorption energy (0.53 eV), in which the O atom is located on top of a Pt atom and the C atom lies above the midpoint between the two adjacent Pt atoms. DME adsorption involves electron transfer (0.1 $e$) from Pt to oxygen atom. The moderate adsorption energy falls in between the physisorbed and chemisorbed states. In addition, the slightly increasing length of the C-O bond implies that DME could not be decomposed into methoxide ($CH_3O^*$) fragments on Pt(100) at the early stage of the dehydrogenation process.

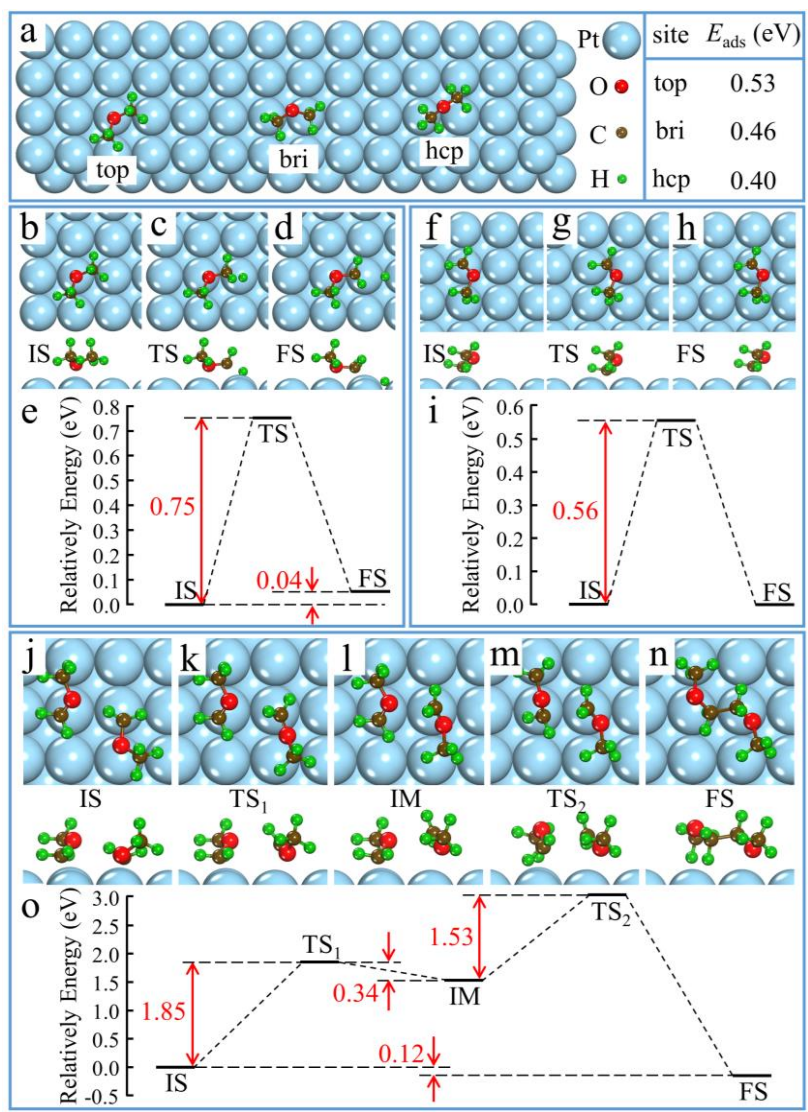

**Figure 12.** (a) Three possible adsorption sites (hcp, top, and bri) for DME adsorption on Pt(100). The adsorption energy for each site is summarized. Optimized structures (top and side views) of the initial state (IS), transition state (TS), intermediate state (IM), and final state (FS) during dehydrogenation (b-d), diffusion (f-h), and coalescence (j-n) processes. Energy profiles for (e) DME dehydrogenation, (i) $CH_3OCH_2^*$ diffusion, and (o) $CH_3OCH_2^*$ rotation-coalescence.

At stage (2), the hydrogen atom of DME closest to the Pt(100) surface tends to be removed. The fully relaxed $CH_3OCH_2^*$ adsorbed on the metal surface is plotted in Fig. 12d. The carbon atom ($C_d$) of the dehydrogenated methyl group moves toward the Pt surface and shows mixed $sp^2$-$sp^3$ hybridization. The length of $C_d$-Pt bond is 2.181 Å, suggesting the covalent interaction between $C_d$ and Pt. The reaction energy ($E_{re}$) is expressed by $E_{re} = E_{FS} - E_{IS}$, where $E_{IS}$ and $E_{FS}$ are the energies of an adsorbed DME molecule and a $CH_3OCH_2^*$ species with one H atom on the surface, respectively. The $E_{re}$ is only 0.04 eV, indicating the dehydrogenation of a DME molecule on Pt(100) is energetically slightly unfavourable. The activation barrier for the dehydrogenation process is 0.75 eV (Fig. 12e), which is much smaller than that of coronene (1.87 eV)[20] and methane (1.77 eV)[65] on the Cu(111) surface.

As mentioned above, the C-O bond will break when DME becomes completely dehydrogenated. In addition, the small carbon-containing species are not stable without H. Thus we only consider the diffusion and coalescence of partially dehydrogenated DME ($CH_3OCH_2^*$) on Pt(100). The relaxed configuration for $CH_3OCH_2^*$ (Fig. 12f) has an adsorption energy of 3.23 eV. Moreover, the relatively low diffusion barrier is 0.56 eV (Fig. 12i), which would be beneficial for the coalescence of $CH_3OCH_2^*$ species. During the coalescence, the system passes through an intermediate state with a rotational degree of freedom. When two similarly charged $CH_3OCH_2^*$ species approach each other, the Coulomb repulsion can serve as a driving force for the rotational movement of $CH_3OCH_2^*$. In the MI (Fig. 12l), one $CH_3OCH_2^*$ rotates by 60° compared to the IS. The calculated rotation barrier is 1.85 eV. With the further approaching of the reactants, the other $CH_3OCH_2^*$ rotates by about 80° with the barrier of 1.53 eV. Figure 12n displays the $C_4O_2H_{10}$ configuration with a covalent C-C bond (1.526 Å) between two adjacent $CH_3OCH_2^*$ species. The formation energy ($E_f$) of $C_4O_2H_{10}$ is defined as $E_f = E_{2(CH_3OCH_2^*)} - E_{C_4O_2H_{10}}$, where $E_{C_4O_2H_{10}}$ is the total energy of $C_4O_2H_{10}$ on Pt(100), respectively. $E_{2(CH_3OCH_2^*)}$ is the energy of two separately adsorbed $CH_3OCH_2^*$ species on the Pt surface. The calculated formation energy is 0.18 eV, suggesting the coalescence is an exothermic process.

However, it should be pointed out that we only clarify the possibility of an alternative reaction pathway for DME on Pt(100) surface. This does not represent a practical method of fabricating the graphether monolayer. A recent advance in the bottom-up fabrication of graphene has demonstrated that the full dehydrogenation occurs at a very late stage with large active $C_xH_y$ species already formed.[66] Therefore, at the initial stage, graphether nucleation could be a continuous $C_xH_yO_z^*$ aggregation and reaction process.

### 3.9 Isoelectronic Analogue

Inspired by the isoelectronic relationship between graphene and *h*-BN, it would be interesting to explore an isoelectronic analogue of graphether where each couple of bonded carbon atoms is replaced by a boron-nitrogen pair. Because the new designed 2D structure has *Pmn*2$_1$ symmetry, we name it as *Pmn*2$_1$-BNO (Fig. 13a). The lattice parameters of *Pmn*2$_1$-BNO are $a = 3.658$ Å and $b = 2.649$ Å, which are slightly larger than those of graphether. The cohesive energy is 8.62 eV/atom, which is comparable with that of monolayer *h*-BN (8.79 eV/atom) calculated at the same level of theory. Furthermore, no imaginary frequency is observed in the whole Brillouin zone (Fig. 13b), implying that the predicted *Pmn*2$_1$-BNO structure is dynamically stable.

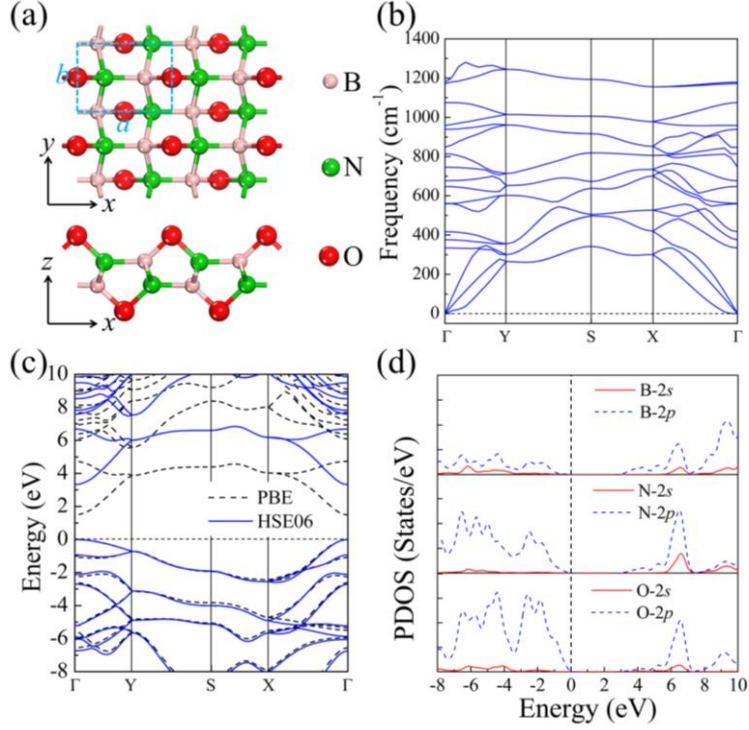

**Figure 13.** (a) Top and side views of the optimized structure of $Pmn2_1$-BNO. (b) Phonon dispersion, (c) band structure, and (d) PDOS of $Pmn2_1$-BNO computed at the HSE06 level.

**Table 4.** Calculated in-plane stiffness ($C_{2D}$), DP constant ($E_{DP}$), effective mass ($m^*$), and carrier mobility ($\mu$) along armchair and zigzag directions for $Pmn2_1$-BNO at 300 K.

| carrier type | $C_{2D}$ (N m$^{-1}$) | $E_{DP}$ (eV) | $m^*$ | $\mu$ (cm$^2$ V$^{-1}$ s$^{-1}$) |
|---|---|---|---|---|
| electron (armchair) | 342.5 | -3.09 | 0.57 | 690 |
| hole (armchair) |  | -0.33 | 2.65 | 102 |
| electron (zigzag) | 202.7 | -8.27 | 0.34 | 657 |
| hole (zigzag) |  | -8.25 | 0.52 | 212 |

As shown in Table 4, the in-plane stiffness along the armchair direction is about 1.8 times larger than that along the zigzag direction. This is because the B-N bond strength along $a$ is stronger than that along $b$ as the B-N bond lengths are 1.554 Å along $a$ and 1.630 Å along $b$. Moreover, the in-plane stiffness of $Pmn2_1$-BNO along the armchair direction (342.5 N m$^{-1}$) is significantly larger than that of monolayer $h$-BN (~280 N m$^{-1}$).[67]

In comparison with $h$-BN which is an insulator with an indirect gap of 5.96 eV,[68]

*Pmn*2$_1$-BNO is a direct gap semiconductor with a moderate band gap of 3.32 eV (Fig. 13c). The PDOS shows that both valence and conduction bands are predominantly composed of 2*p* orbitals of B, N and O atoms (Fig. 13d). This suggests that oxidation can affect the electronic properties and reduce the band gap of *h*-BN.

In Table 4, the anisotropy of the hole mobility is stronger than that of the electron mobility. The hole mobility along the zigzag direction is about 2 times larger than that along the armchair direction, leading to the zigzag direction more favorable for the hole conduction. However, the electron mobility shows isotropy behavior and is higher than that of monolayer *h*-BN (487 cm$^2$ V$^{-1}$ s$^{-1}$).[12] Furthermore, the large difference between electron and hole may facilitate the migration and separation of photogenerated electron-hole pairs.

## 4 Conclusions

In summary, inspired by the hyperconjugative interactions in DME, we have theoretically designed a stable graphether monolayer with a direct wide-band-gap of 2.39 eV. The direct band gap feature is robust against external perturbations such as uniaxial strain, biaxial strain, and layer stacking. Moreover, graphether exhibits strong light absorption in the ultraviolet region. The hyperconjugation interaction is identified in graphether, which leads to the high in-plane stiffness (459.8 N m$^{-1}$) and electron mobility (1.8×10$^3$ cm$^2$ V$^{-1}$) along the armchair direction. We have demonstrated that Pt(100) would be a potential substrate for the bottom-up synthesis of graphether using DME precursors. Finally, its isoelectronic analogue, *Pmn*2$_1$-BNO monolayer, has been proposed, which is also a direct wide-band-gap semiconductor with attractive mechanical and electrical properties. All these findings suggest that graphether is a promising candidate for next-generation nanoelectronic devices. We expect our results will promote the experimental fabrication of graphether and attract attentions on designing 2D compounds with hyperconjugation effects.

## Conflicts of interest

There are no conflicts to declare.

## Acknowledgments

This work is supported by the National Natural Science Foundation of China (Grant Nos. 61974068 and 11704198), the Jiangsu Provincial Government Scholarship Program, and the Summit of the Six Top Talents Program of Jiangsu Province (Grant No. 2015-XCL-022).

# Supplementary Information

**Graphether：A Two-Dimensional Oxocarbon as a Direct Wide-Gap Semiconductor with High Mechanical and Electrical Performances**

Gui-Lin Zhu[1∥], Xiao-Juan Ye[1,2∥], and Chun-Sheng Liu[1,2*]

[1]Key Laboratory of Radio Frequency and Micro-Nano Electronics of Jiangsu Province, College of Electronic and Optical Engineering, Nanjing University of Posts and Telecommunications, Nanjing 210023, China

[2]School of Engineering, University of British Columbia, Kelowna, BC V1V 1V7, Canada

* To whom correspondences should be addressed, E-mail: csliu@njupt.edu.cn
∥ The first two authors contributed equally to this work.

**Table S1.** Several schemes for band gap opening in graphene.

| Scheme | | | Maximum Band Gap (eV) |
|---|---|---|---|
| spatial confinement | width ~15 nm graphene nanoribbon | | 0.2 (Ref. 1) |
| | width N=7 graphene nanoribbon | | 1.6 (Ref. 2) |
| sublattice-symmetry breaking | strain engineering | uniaxial strain | 0.486 (Ref. 3) |
| | | anisotropic biaxial strain | 1.0 (Ref. 4) |
| | | shear strain | 0.4 (Ref. 4), 0.95 (Ref. 5) |
| | doping | B | 0.14 (Refs. 6,7) |
| | | N | 0.14 (Refs. 6,7) |
| | | O | 0.50 (Ref. 7) |
| | | P | 0.67 (Ref. 8) |
| | | S | 0.57 (Ref. 8) |
| | | Ga | 0.6 (Ref. 9) |
| | | Ge | 0.4 (Ref. 9) |
| | | As | 1.3 (Ref. 9) |
| | | Se | 0.8 (Ref. 9) |
| | graphene antidot lattice | | 1.55 (Ref. 10) |
| surface functionalization | fully hydrogenated | | 3.7 (Ref. 11) |
| | fully fluorinated | | 3.48 (Ref. 12) |
| | half hydrogenated | | 0.43 (Ref. 12) |
| | half hydrogenated and half fluorinated | | 3.70 (Ref. 12) |
| | fully oxidized | | 4.0 (Ref. 13) |
| heterostructure | graphene/h-BN | | 0.053 (Ref. 14), 0.16 (Ref. 15) |
| | graphene/SiC | | 0.26 (Ref. 16) |
| | graphene/fully hydroxylated $SiO_2$ | | 0.023 (Ref. 17) |
| | graphene/O-terminated $SiO_2$ | | 0.044 (Ref. 17) |
| electrically tunable band gap in multilayer graphene | bilayer | | 0.1 (Ref. 18), 0.25 (Ref. 19) |
| | trilayer | | 0.12 (Ref. 20) |

## Computational details of the carrier mobility and in-plane stiffness

We estimate the carrier mobility for graphether at room temperature (300 K) by using the equation $\mu_x \approx \dfrac{e\hbar^3 \left(\dfrac{5C_x + 3C_y}{8}\right)}{k_B T m_x^{\frac{3}{2}} m_y^{\frac{1}{2}} \left(\dfrac{9E_{1x}^2 + 7E_{1x}E_{1y} + 4E_{1y}^2}{20}\right)}$, where $m_x$ is the effective mass in the transport direction and $m_y$ is the effective mass in another direction. The term $E_1$ defined by $E_1 = \dfrac{\Delta V}{(\Delta l / l_0)}$ represents the DP constant of the valence-band minimum (VBM) for hole or the conduction-band maximum (CBM) for electron along the transport direction. Here, $\Delta V$ denotes the energy change of VBM or CBM when graphether is compressed or dilated from the equilibrium $l_0$ by a distance of $\Delta l$. The term $C_x$ ($C_y$) is the in-plane stiffness of the longitudinal strain in the $x$ ($y$) direction, which can be derived from $\dfrac{\Delta E}{S_0} = C_x \dfrac{(\Delta l / l_0)^2}{2}$. $\Delta E$ denotes the total energy difference under each strain, and $S_0$ is the lattice area of pristine graphether. We use $\Delta l / l_0$ ranging from -0.5% to 0.5% to fit the values of $C_{2D}$ and $E_1$ (Fig. S1).

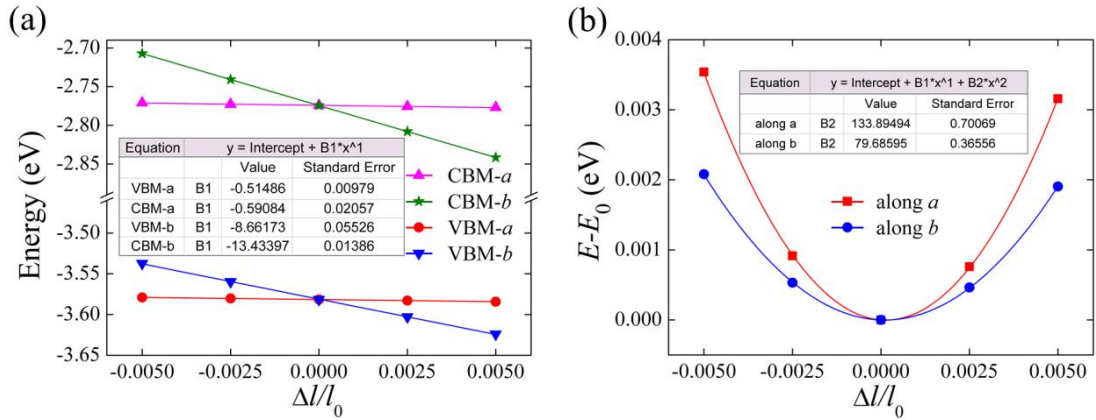

**Fig. S1** (a) DP constants and (b) in-plane stiffness of graphether along $x$ and $y$ directions at the HSE06 level of theory.

**Table S2.** Lattice parameter, interlayer distance ($d_z$), binding energy ($E_b$), and band gap for bilayer graphether.

| stacking pattern | lattice parameters (Å) | | $d_z$ (Å) | $E_b$ (meV Å$^{-2}$) | band gap (eV) |
|---|---|---|---|---|---|
| | a | b | | | |
| monolayer | 7.229 | 5.155 | - | - | 2.39 |
| AA | 7.233 | 5.152 | 2.003 | -33.44 | 2.27 |
| AB | 7.227 | 5.154 | 3.451 | -9.42 | 2.31 |
| AC | 7.233 | 5.150 | 2.150 | -27.75 | 2.27 |
| AD | 7.227 | 5.154 | 3.516 | -9.12 | 2.32 |

**Table S3.** Structural informations of monolayer graphether and *Pmn*2$_1$-BNO.

| Structure | Space Group | Lattice Parameters | Wyckoff Positions (fractional) | | | |
|---|---|---|---|---|---|---|
| | | | atoms | x | y | z |
| Graphether | *Pmmn* | a=3.6144 Å | C(4f) | 0.2881 | 0 | 0.5180 |
| | | b=2.5775 Å | O(2a) | 0 | 0 | 0.5570 |
| | | c=25.0000 Å | | | | |
| | | α=β=γ=90° | | | | |
| *Pmn*2$_1$-BNO | *Pmn*2$_1$ | a=2.6493 Å | B(2a) | 0 | 0.4811 | 0.7062 |
| | | b=25.0000 Å | N(2a) | 0 | 0.4825 | 0.2816 |
| | | c=3.6579 Å | O(2a) | 0 | 0.4414 | 0.9897 |
| | | α=β=γ=90° | | | | |